\documentclass[a4paper]{article}

\usepackage{microtype}
\usepackage[usenames]{xcolor}
\usepackage{bbm}
\usepackage{lineno}
\usepackage{hyperref}
\usepackage{amsthm}
\usepackage{amsmath}
\usepackage{graphicx}
\usepackage{amssymb}
\usepackage{url}
\usepackage{amsopn}
\usepackage{authblk}

\newcommand{\changed}[1]{{#1}}

\newcommand{\remove}[1]{{}}

\newcommand{\tcomplex}{\ensuremath{\mathbbm{T}}}

\newcommand{\fcomplex}{\ensuremath{\mathbbm{X}}}

\newcommand{\ospoly}{\ensuremath{\mathbbm{Y}}}

\theoremstyle{plain}
\newtheorem{theorem}{Theorem}
\newtheorem{lemma}[theorem]{Lemma}

\newtheorem{proposition}[theorem]{Proposition}
\theoremstyle{definition}

\theoremstyle{remark}

\title{A Proof of the Orbit Conjecture for Flipping Edge-Labelled Triangulations}
\date{}

\author[1]{Anna Lubiw}
\author[2]{Zuzana Mas\'arov\'a}
\author[2]{Uli Wagner}
\affil[1]{School of Computer Science, University of Waterloo\\
 Waterloo, ON, Canada, N2L 3G1\\
  \texttt{alubiw@uwaterloo.ca}}
\affil[2]{IST Austria\\
  Am Campus 1, 3400 Klosterneuburg, Austria\\
  \texttt{zuzana.masarova@ist.ac.at, uli@ist.ac.at}}
  
\begin{document}

\maketitle

\begin{abstract}
Given a triangulation of a point set in the plane, a \emph{flip} deletes an edge $e$ whose removal leaves a convex quadrilateral, and replaces $e$ by the opposite diagonal of the quadrilateral. 
It is well known that any triangulation of a point set can be reconfigured to any other triangulation by some sequence of flips.
We explore this question in the setting where each edge of a triangulation has a label, and a flip transfers the label of the removed edge to the new edge.
It is not true that every labelled triangulation of a point set can be reconfigured to every other labelled triangulation via a sequence of flips, but we characterize when this is possible.
There is an obvious necessary condition: for each label $l$,  
if edge $e$ has label $l$ in the first triangulation and edge $f$ has label $l$ in the second triangulation,
then there must be some sequence of flips that moves label $l$ from $e$ to $f$, ignoring all other labels.   
Bose, Lubiw, Pathak and Verdonschot 
formulated the \emph{Orbit Conjecture}, which states
that this necessary condition is also sufficient, i.e.~that \emph{all} labels can be simultaneously mapped to their destination if and only if \emph{each} label individually can be mapped to its destination.
We prove this conjecture. 
Furthermore, we give a polynomial-time algorithm to find a sequence of flips 
to reconfigure one labelled triangulation to another, if such a sequence exists, and
we prove an upper bound of $O(n^7)$ 
on the length of the flip sequence.  

Our proof uses the topological 
result 
that 
the sets of pairwise non-crossing edges on a planar point set form 
a simplicial complex that 
is homeomorphic to a high-dimensional ball 
(this follows from a result of Orden and Santos; we 
give a different proof based 
on a shelling argument). 
The dual cell complex of this simplicial ball, 
called the
\emph{flip complex}, has the usual flip graph as its $1$-skeleton.  We use properties 
of the $2$-skeleton of the flip complex 
to prove the Orbit Conjecture.


\end{abstract}

 \section{Introduction}

The flip operation is fundamental to the study of triangulations of point sets in the plane.
A flip removes one edge and replaces it by the opposite diagonal of the resulting quadrilateral, so long as that quadrilateral is convex.  
Lawson~\cite{Lawson-72} proved the foundational result that any triangulation can be transformed into any other triangulation of the same point set via a sequence of flips.
His second proof of this result~\cite{Lawson-77} used the approach that is more widely known---showing that any triangulation can be flipped to the Delaunay triangulation, which then acts as a ``hub'' through which we can 
flip any triangulation to any other.

The result that any triangulation can be flipped to any other is captured succinctly by saying that the 
\emph{flip graph} is connected, where 
the \emph{flip graph} has a vertex for each triangulation of the given point set, and an edge when two triangulations differ by one flip. 
The special case of a point set in convex position has been very thoroughly studied.  In this case triangulations correspond to binary trees, and a flip corresponds to a rotation.
The flip graph in this case is the 1-skeleton of a polyhedron called the \emph{associahedron}.

The use of flips to reconfigure triangulations is relevant to the study of associahedra~\cite{STT88} and mixing~\cite{molloy1999mixing}. 
Flips are also important in practice for mesh generation and 
for finding triangulations that optimize certain quality measures~\cite{Bern-Eppstein,Edelsbrunner}.
The survey by Bose and Hurtado~\cite{BH09} discusses these and many other aspects of flips.  

Despite the extensive work on flips, it is only recently that the question of where edges go under flip operations has been investigated. 
This can be formalized by attaching a label to each edge in a triangulation.
Throughout, we fix a set $P$ of $n$ points in general position, and we identify triangulations with their edge sets (i.e., a triangulation of $P$ is a maximal set $T$ of pairwise non-crossing edges spanned by $P$).
A \emph{labelled triangulation} $\cal T$ of $P$ is a pair $(T, \ell)$ where $T$ is a triangulation of $P$  and $\ell$ is a \emph{labelling function} that maps the edges of $T$ one-to-one onto the labels $1, 2, \ldots, t_P$.  Here $t_P$ is the number of edges in any triangulation of $P$.  
When we perform a flip  operation on $\cal T$, 
 the label of the removed edge is transferred to the new edge.

We can now capture ``where an edge goes'' under flip operations.  
We say that edges $e$ and $f$ lie in the same \emph{orbit} if we can attach label $l$ to $e$ in some triangulation and apply some sequence of flips to arrive at a triangulation in which edge $f$ has label $l$.
The orbits are exactly the connected components of a graph that 
Eppstein~\cite{Eppstein} called the \emph{quadrilateral graph}---this graph has a vertex for every one of the possible $n \choose 2$ edges formed by point set $P$, 
with $e$ and $f$ being adjacent if they cross
and their four endpoints form a convex quadrilateral that is empty of other points.  In particular, this implies that there is a polynomial-time algorithm to find the orbits.
The orbits can be very different depending on $P$.
For a point set in convex position, all the non-convex hull edges are in a single orbit~\cite{bose2013flipping}, but at the other extreme, a point set with no empty convex pentagon has 
the property that 
in any triangulation, the edges are all in distinct orbits~\cite{Eppstein}.

Orbits tell us where each individual edge label can go, but not how they combine.
The main question we address in this paper is: when is there a sequence of flips to reconfigure one labelled triangulation of point set $P$ to another labelled triangulation of $P$?
A necessary condition is that,  for each label $l$, 
the edges with label $l$ in the two triangulations
must lie in the same orbit.
Bose et al.~\cite{bose2013flipping} conjectured that this condition is also sufficient.
As our main result we prove this ``Orbit Conjecture,''  and strengthen it by providing a polynomial-time algorithm and a bound on the length of the flip sequence.

\begin{theorem}[Orbit Theorem] 
\label{thm:orbit}
 Given two edge-labelled triangulations ${\cal T}_1$ and ${\cal T}_2$ of a point set,
there is a flip sequence that transforms one into the other if and
only if for every label $l$, the edges of ${\cal T}_1$ and ${\cal T}_2$ having label
$l$ belong to the same orbit.
Furthermore, there is a polynomial-time algorithm that tests whether the condition 
is satisfied, and if it is, computes a flip sequence of length $O(n^7)$ to transform ${\cal T}_1$ to ${\cal T}_2$. 
\end{theorem}


The orbit theorem is stated for triangulations ${\cal T}_1$ and ${\cal T}_2$ that may have different edge sets, but---since we know how to use flips to change the edge set---the crux of the matter is the special case where the two triangulations have the same edge set $T$ but different label functions $\ell_1$ and $\ell_2$.
In other words, we are given a  permutation 
of the edge labels of a triangulation, and we seek a flip sequence to realize the permutation. 
Furthermore, since every permutation is a composition of transpositions, 
we concentrate first on finding a flip sequence to transpose (or ``swap'')  two labels.
This idea of reducing the problem to the case of swaps appears
in~\cite{bose2013flipping}.

\begin{figure}
\centering
\includegraphics[width=\linewidth]{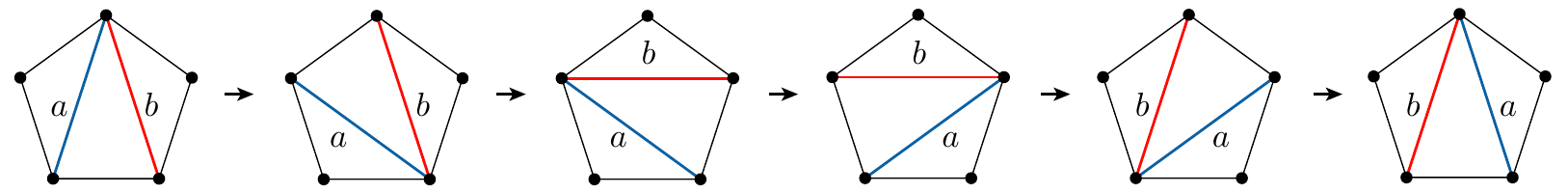}
\caption{Five flips swap the edge labels ($a$ and $b$) of two diagonals of a convex pentagon.  In the flip graph these five flips form a 5-cycle.}
\label{fig:basic-pentagon-swap}
\end{figure}

One insight to be gained from previous work is that empty convex pentagons in the point set seem to be crucial for 
swapping edge labels.
Certainly, an empty convex pentagon provides a label swap---Figure~\ref{fig:basic-pentagon-swap} shows how the edge labels of two diagonals of an empty convex pentagon can be swapped by a sequence of five flips.  
In the other direction, the special cases of the orbit theorem that were proved by Bose et al.~\cite{bose2013flipping} for convex and spiral polygons involved moving pairs of labels into empty convex pentagons and swapping them there. 
Furthermore, Eppstein~\cite{Eppstein} showed that in a triangulation of a point set with no empty convex pentagons, no permutations of edge labels are possible via flips.

The foundation of our proof is to make this intuition about empty convex pentagons rigorous.  
In particular, we show that 
the only elementary operation that is needed 
for label permutation 
is to transpose two labels by moving them into an empty convex pentagon and swapping them there.
More formally, given a labelled triangulation $\cal T= (T,\ell)$, an \emph{elementary swap} of edges $e$ and $f$ in $T$ is a transposition of the labels of $e$ and $f$ that is accomplished as follows: perform a sequence, $\sigma$, of flips on $\cal T$ to get to a triangulation $\cal T'$ 
in which the labels $\ell(e)$ and $\ell(f)$ are attached to the two diagonals of an empty convex pentagon;
then perform the 5-flip sequence, $\pi$, that transposes these two labels; then perform the sequence $\sigma^{-1}$.  We say that the sequence $\sigma \pi \sigma^{-1}$ \emph{realizes} the elementary swap.
Observe that the effect of $\sigma \pi \sigma^{-1}$ on $\cal T$ is to transpose the labels of $e$ and $f$ while leaving all other labels unchanged.  
We will prove that an elementary swap can always be realized by a flip sequence of length $O(n^6)$, and furthermore, that such a sequence can be found in polynomial time.

One of our main results is the following, from which the Orbit Theorem can readily be derived:

\begin{theorem}
\label{thm:swap}
In a labelled triangulation $\cal T$, two edges are in the same orbit if and only if there is an elementary swap between them.
\end{theorem} 

In order to prove Theorem~\ref{thm:swap}, we use the following key result:

\begin{theorem}[Elementary Swap Theorem]
\label{thm:elementary-swap-2}
Given a labelled triangulation $\cal T$, 
any permutation of the labels that can be realized by a sequence of flips can be realized by a sequence of elementary swaps.
\end{theorem}

This theorem is proved using topological properties of the \emph{flip complex}, whose $1$-skeleton is the flip graph.  A result of Orden and Santos~\cite{Orden:The-polytope-of-non-crossing-graphs-on-a-planar-2005} can be used to show 
that the flip complex has the topology of a high-dimensional ball\footnote{Technically speaking, the flip complex is homotopy equivalent to a ball.}. We give an alternate proof of this.
We use the $2$-skeleton of the flip complex, and show that its $2$-cells correspond to cycles in the flip graph of two types:
quadrilaterals, which do not permute labels;  and pentagons, which correspond precisely to the 5-cycles of flips shown in Figure~\ref{fig:basic-pentagon-swap}.
Then we prove the Elementary Swap Theorem by translating it into a result about decomposing closed walks in the flip graph into simpler \emph{elementary walks}.

Although there is a rich literature on associahedra and on 
cell complexes associated with triangulations of point sets, we are not aware of any previous combinatorial results on triangulations that require topological proofs, as our proof of the Orbit Theorem seems to.

We now briefly describe the rest of our method after 
the Elementary Swap Theorem is established. 
In order to prove Theorem~\ref{thm:swap}, we need one more ingredient about the structure of elementary swaps: we will show that any sequence of elementary swaps that moves the label of edge $e$ to edge $f$ can be ``completed''  to get the label of $f$ back to $e$, and that, in fact, the 
resulting sequence provides an elementary swap of $e$ and $f$.

The high-level idea of our proof of Theorem~\ref{thm:swap} is then as follows:
From our hypothesis that two edges $e$ and $f$ lie in the same orbit, we show that there is a sequence of flips that permutes the labels of triangulation $\cal T$, taking the label of $e$ to $f$.   The Elementary Swap Theorem then gives us a sequence of elementary swaps to do the same (this is the significant step of the proof).  Finally, from the structure of elementary swaps we can then find an elementary swap of $e$ and $f$.

Our paper is organized as follows.  
In Section~\ref{sec:topology} we prove the Elementary Swap Theorem using topological methods.  In Section~\ref{sec:bounds} we prove the properties of elementary swaps that were mentioned above. 
In top-down fashion, we begin in Section~\ref{sec:reductions} by expanding on the high-level ideas, and proving the Orbit Theorem assuming the results in the later sections.

\subsection{Background}
\label{sec:background}

The diameter of the flip graph of a point set gives the 
worst-case number of flips required to reconfigure one triangulation to another.
For unlabelled triangulations, the diameter of the flip graph 
is known to be $\Theta(n^2)$, with the upper bound proved by Lawson~\cite{Lawson-72} and the lower bound  proved by Hurtado~et al.~\cite{HNU99}.
For the special case of points in convex position, there is an exact bound of $2n-10$~\cite{STT88,Pournin13}. 
The problem of finding the distance in the flip graph between two given triangulations of a point set is NP-hard~\cite{lubiw2015flip}, and even APX-hard~\cite{pilz2014flip}.  
\changed{It has recently been shown to be fixed-parameter tractable~\cite{kanj-2017}.}
The problem remains NP-hard for triangulations of a polygon~\cite{aichholzer2015flip}, but the complexity status is open for the case of points in convex position. 
For further results on flips, see the 
survey by Bose and Hurtado~\cite{BH09}.

The \emph{labelled flip graph} of a point set has a vertex for every labelled triangulation of the point set and an edge when two labelled triangulations differ by a flip.
Bose et al.~\cite{bose2013flipping} 
formulated the Orbit Conjecture and proved it 
for the special case of triangulations of any convex polygon, 
showing that the labelled flip graph has a single connected component (ignoring convex hull edges, which cannot flip), and  giving a tight bound of $\Theta(n \log n)$ on its diameter. 
Araujo-Pardo et al.~\cite{AHOS14} independently proved the Orbit Conjecture for convex polygons, and introduced ``colorful associahedra'' which generalize associahedra to the setting of labelled (or coloured) triangulations.
Bose~et~al.~also proved the Orbit Conjecture for spiral polygons.  In this case the labelled flip graph may be disconnected but each connected component has diameter $O(n^2)$, which is a tight bound.

The best known lower bound on the diameter of a connected component of the labelled flip graph for a point set is $\Omega(n^3)$~\cite{bose2013flipping}.  There is a large gap between this lower bound and our upper bound of $O(n^7)$.

The Orbit Theorem holds for combinatorial triangulations~\cite{bose2013flipping}, and for pseudotriangulations~\cite{bose2015flips}.  In both these cases there is a single orbit, 
so the labelled flip graph is connected.
There are also some related results using variants of the flip operation, for example, 
Cano~et al.~\cite{JDHU13} reconfigured edge-labelled non-maximal plane graphs by ``rotating'' edges around one of their endpoints; again there is a single orbit. 
A related result where there are multiple orbits is an analogue of the Orbit Theorem for labelled (or ``ordered'') bases of a matroid---one labelled basis can be turned into another labelled basis via basis exchange steps if and only if elements with the same label lie in the same connected component of the matroid~\cite{lubiw-pathak-matroids}.

For more general problems of reconfiguring one structure to another via elementary steps, see~\cite{IDHPSUU11,Heu13}.  


 \subsection{Preliminaries and Definitions}
\label{sec:preliminaries}

Most definitions were given above,
but we fill in a few missing details.
Throughout, we assume a set of $n$ point in general position in the plane.
A point set determines $n \choose 2$ \emph{edges} which are the line segments between pairs of points.  Two edges \emph{cross} if they intersect in a point that is interior to at least one of the two edges.
An \emph{empty convex k-gon} is a subset of $k$ points that forms a convex polygon with no point of $P$ in its interior.  A \emph{diagonal} of a convex polygon is an edge joining two points that are not consecutive on the polygon boundary.

Several times in our proofs we will use the result that if two unlabelled triangulations of the same point set have a subset, $S$, of \emph{constrained} edges in common, then there is a sequence of flips that transforms one triangulation into the other, without ever flipping any edge of $S$, i.e.~the edges in $S$ remain fixed throughout the flip sequence.   
This was first proved by Dyn et al.~\cite{DGR93}, and can alternatively be proved using constrained Delaunay triangulations~\cite{Bern-Eppstein}. 



\section{Proof of the Orbit Theorem}
\label{sec:reductions}

In this section we prove the Orbit Theorem assuming the Elementary Swap Theorem (Theorem~\ref{thm:elementary-swap-2}, proved in Section~\ref{sec:topology}), and assuming the following two results on elementary swaps.
The first result shows that 
every elementary swap can be realized by a relatively short flip sequence that can be found efficiently,
and the second result gives us a way to combine elementary swaps so that, after moving $e$'s label to $f$, we can get $f$'s label back to $e$.
These lemmas will be proved in Section~\ref{sec:bounds}. 

\begin{lemma}
\label{lemma:elem-swap}
If there is an elementary swap between two edges in a triangulation $\cal T$ then there is a flip sequence of length $O(n^6)$ to realize the elementary swap, and, furthermore, this sequence can be found in polynomial time.
\end{lemma}

\begin{lemma} 
\label{lemma:elem-swap-seq}
Let $\cal T$ be a labelled triangulation containing two edges $e$ and $f$.   If there is a sequence of elementary swaps on $\cal T$ that takes the label of edge $e$ to edge $f$, then there is 
an elementary swap of $e$ and $f$ in $\cal T$.
\end{lemma}

As we show in Section~\ref{sec:bounds}, 
a simple group-theoretic argument suffices to prove a weaker version of Lemma~\ref{lemma:elem-swap-seq}, namely, that under the stated assumptions, there is a sequence of elementary swaps exchanging the labels of $e$ and $f$ in $\cal T$.  
Proving the stronger version, which we need for our bounds on the length of flip sequences, requires using the properties of elementary swaps.

We prove the Orbit Theorem in stages, first Theorem~\ref{thm:swap} (the case of swapping two labels in a triangulation),
then the more general case of permuting edge labels in a triangulation, and finally the full result.

\begin{proof}[Proof of Theorem~\ref{thm:swap}] 
The ``if'' direction is clear, so we address the ``only if'' direction.
Suppose that ${\cal T} = (T, \ell)$ is the given edge-labelled triangulation and that 
$e$ and $f$ are edges of $T$ that are in the same orbit.  
Then
there is a sequence of flips that changes ${\cal T}$ to an edge-labelled triangulation ${\cal T}' = (T', \ell')$ where $T'$  contains $f$ and $\ell'(f) = \ell(e)$.  
We now apply the result that any constrained triangulation of a point set can be flipped to any other.
Fix edge $f$ and flip $T'$ to $T$.  
Applying the same flip sequence to the labelled triangulation ${\cal T}'$ yields an edge-labelling of triangulation $T$ in which edge $f$ has the label $\ell(e)$.  Thus we have a sequence of flips that permutes the labels of $\cal T$ and moves the label of $e$ to $f$.

By the Elementary Swap Theorem (Theorem~\ref{thm:elementary-swap-2}) there is a sequence of elementary swaps whose effect is to move the label of edge $e$ to edge $f$.  
By Lemma~\ref{lemma:elem-swap-seq} there is an elementary swap of $e$ and $f$ in $\cal T$. 
 \end{proof}


\begin{theorem}[Edge Label Permutation Theorem]
\label{thm:strong-permutation}
Let $T$ be a triangulation of a point set with two edge-labellings $\ell_1$ and $\ell_2$ such that for each label $l$, the edge with label $l$ in $\ell_1$ and the edge with label $l$ in $\ell_2$ are in the same orbit.  Then there is a sequence of $O(n)$ elementary swaps to transform the first labelling to the second.  
Such a sequence can be realized via a sequence of $O(n^7)$ flips, which can be found in polynomial time.
\end{theorem}
\begin{proof}
The idea is to effect the permutation as a sequence of swaps. 
If every edge has the same label in $\ell_1$ and $\ell_2$ we are done.   
So consider a label $l$ that is attached to a different edge in $\ell_1$ and in $\ell_2$.
Suppose $\ell_1(e) = l$ and $\ell_2(f) = l$, with $e \ne f$.
By hypothesis, $e$ and $f$ are in the same orbit.  
By Theorem~\ref{thm:swap} there is an  elementary swap of $e$ and $f$ in $(T, \ell_1)$ which results in a new labelling $\ell_1'$ that matches $\ell_2$ in one more edge (namely the edge $f$) and still has the property that for every label $l$, the edge with label $l$ in  $\ell_1'$ and the edge with label $l$ in $\ell_2$ are in the same orbit.  
Thus we can continue this process until all edge labels match those of $\ell_2$.
In total we use $O(n)$ elementary swaps. These can be realized via a sequence of $O(n^7)$ flips by Lemma~\ref{lemma:elem-swap}.   Furthermore, the sequence can be found in polynomial time.  
\end{proof}


We can now prove the Orbit Theorem.

\begin{proof}[Proof of Theorem~\ref{thm:orbit}]
The necessity of the condition is clear, and we can test it in polynomial time by finding all the orbits, so we address sufficiency.
The idea
 is to reconfigure ${\cal T}_1$ to have the same underlying unlabelled triangulation as ${\cal T}_2$ and then apply the previous theorem.  The details are as follows. 
Let ${\cal T}_1 = (T_1, \ell_1)$ and ${\cal T}_2 = (T_2, \ell_2)$. 
There is a sequence $\sigma$ of $O(n^2)$ flips to reconfigure the unlabelled triangulation $T_1$ to $T_2$, and $\sigma$ can be found in polynomial time.
Applying $\sigma$ to the labelled triangulation ${\cal T}_1$ yields a labelled triangulation ${\cal T}_3 = (T_2, \ell_3)$.  
Note that for every label $l$, the edges of ${\cal T}_1$ and ${\cal T}_3$ having label $l$ belong to the same orbit.
This is because flips preserve orbits (by definition of orbits).
Thus by Theorem~\ref{thm:strong-permutation} there is a flip sequence $\tau$ that reconfigures ${\cal T}_3$ to ${\cal T}_2$, and this flip sequence can be found in polynomial time and has length $O(n^7)$.
The concatenation of the two flip sequences, $\sigma \tau$, reconfigures ${\cal T}_1$ to ${\cal T}_2$, has length $O(n^7)$, and can be found in polynomial time.  
\end{proof}


\section{Proof of the Elementary Swap Theorem}
\label{sec:topology}

As mentioned in the introduction, we prove the Elementary Swap Theorem using topological properties of the \emph{flip complex}, whose 1-skeleton (i.e.~vertices and edges) is the flip graph. 
In fact, we will only need the 2-cells of the flip complex, not any higher-dimensional structure. 
We will show that 2-cells of the flip complex correspond to $4$- and $5$-cycles in the flip graph.

The basic idea is as follows.  We will translate the Elementary Swap Theorem to a statement about walks in the flip graph.  
The hypothesis of the Elementary Swap Theorem is that we have a sequence of flips that permutes the edge labels of a triangulation $T$.  In the flip graph, this sequence corresponds to a closed walk $w$ that starts and ends at triangulation $T$.   Our main topological result 
is that the flip complex has a trivial fundamental group, which will imply that such a closed walk $w$ can be decomposed into simpler \emph{elementary walks}.  Each elementary walk starts at $T$, traces a path in the flip graph, then traverses the edges of a 2-cell, then retraces the path back to $T$. 
The edge-label permutation induced by an elementary walk depends on the 2-cell.
If the 2-cell is a $4$-cycle, the permutation is the identity; and if the 2-cell is a $5$-cycle, then the permutation is a transposition, and the elementary walk corresponds to an elementary swap.  Altogether, this implies that the permutation induced by the closed walk $w$ can be expressed as a composition of elementary swaps, which proves the Elementary Swap Theorem.

%
%

\begin{figure}
\begin{centering}
\includegraphics[width=.9\linewidth]{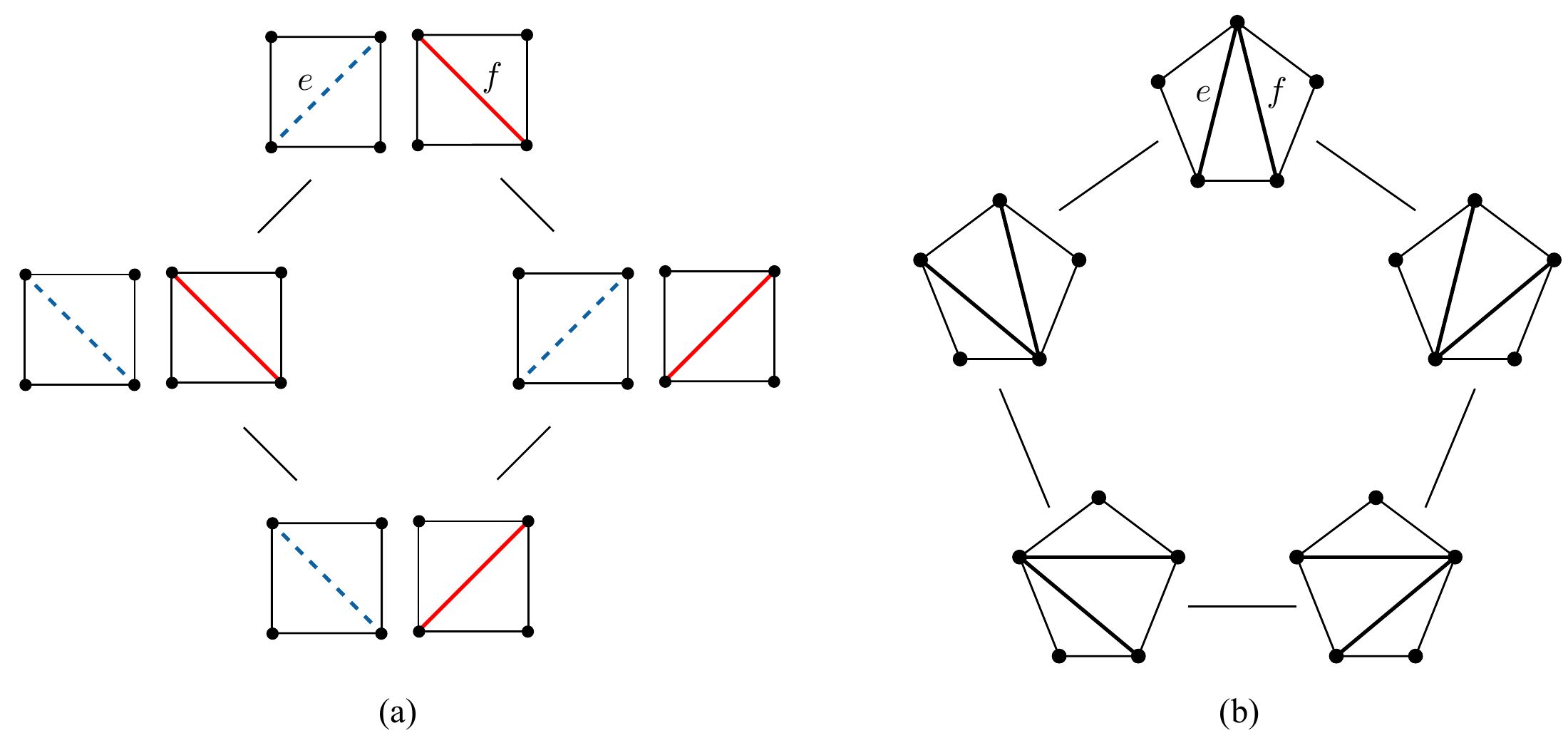}
\par\end{centering}
\protect\caption{(a) Triangulations that differ in the diagonals of two internally
disjoint quadrilaterals form an \emph{elementary $4$-cycle} in the flip graph.  The cycle does not permute the labels (shown as red and blue).  (b)  Triangulations that differ in the diagonals of a convex pentagon form an \emph{elementary 5-cycle} in the flip graph. 
This cycle permutes labels as shown in Figure~\ref{fig:basic-pentagon-swap}.}
\label{4-5cycle}
\end{figure}

Before stating our main topological theorem, we first define the special cycles that will be shown to correspond to $2$-cells of the flip complex. 
In the same way that an edge of the flip complex corresponds to two triangulations that differ on one edge, every 2-cell of the flip complex corresponds to a set of triangulations that differ on two edges.  
Define an \emph{elementary $4$-cycle} to be a cycle of the flip graph obtained in the following way.  Take a triangulation $T$ and two edges $e,f \in T$ whose removal leaves two internally disjoint convex quadrilaterals in $T$.  Each quadrilateral can be triangulated in two ways, which results in four triangulations that contain  $F:=T\setminus \{e,f\}$.  These four triangulations form a $4$-cycle in the flip graph, as shown in Figure~\ref{4-5cycle}(a).  Observe that a traversal of the cycle corresponds to a  sequence of flips that returns edge-labels to their original positions. 

Define an \emph{elementary $5$-cycle} to be a cycle of the flip graph obtained in the following way.  Take a triangulation $T$ and two edges $e,f \in T$ whose removal leaves a convex pentagon in $T$.  There are five triangulations that contain  $F:=T\setminus \{e,f\}$, and they form a $5$-cycle in the flip graph, as shown in Figure~\ref{4-5cycle}(b). Observe that the sequence of flips around such a cycle permutes labels of $e$ and $f$ as shown in Figure~\ref{fig:basic-pentagon-swap}.

As a side remark, note that it can be shown that, in fact, any cycle in the flip graph of length less than 6 is an elementary 4- or 5-cycle. However, we will not need this in what follows.

Our main topological theorem is the following.

\begin{theorem}
\label{thm:flip-complex}
Let $P$ be a set of $n$ points in general position in the plane.
There is a high-dimensional cell complex $\fcomplex=\fcomplex(P)$, which we call the \emph{flip complex}, 
such that:
\begin{enumerate}
\item The 1-skeleton of $\fcomplex$ is the flip graph of $P$;
\item There is a one-to-one correspondence between the 2-cells of $\fcomplex$ and the elementary 4-cycles and elementary 5-cycles of the flip graph of $P$;
\item $\fcomplex$ has the topology of (i.e.,~is homotopy equivalent to) a high-dimensional ball; therefore its \emph{fundamental group}, $\pi_1(\fcomplex)$, is trivial.
\end{enumerate}
\end{theorem}

In what follows, we will use a number of notions from combinatorial topology; some of these 
we will recall along the way, but others we will only describe informally or leave undefined 
and instead refer the reader to standard textbooks for further background (in particular, we refer 
the reader to \cite[Appendix~4.7]{Bjorner:Oriented-matroids-1999} and \cite{Hudson:Piecewise-linear-topology-1969} for background on \emph{regular cell complexes}, \emph{shellability}, and \emph{piecewise linear balls and spheres}, to \cite{Stillwell:Classical-topology-and-combinatorial-group-1993} for background on the fundamental group of cell complexes, and to 
\cite{Hudson:Piecewise-linear-topology-1969,Munkres:Elements-of-algebraic-topology-1984}
for background on \emph{dual complexes}; we will provide more detailed references for specific 
results below).

Theorem~\ref{thm:flip-complex} follows from a result of Orden and Santos~\cite{Orden:The-polytope-of-non-crossing-graphs-on-a-planar-2005}; we are grateful to F.~Santos for bringing this reference to our attention. In fact, Orden and Santos show something stronger: There exists a simple polytope $\ospoly=\ospoly(P)$ and a face $F$ of $\ospoly$ such that $\fcomplex$ can be taken to be the complement of the star of $F$ in $\ospoly$.

Before becoming aware of the work of Orden and Santos, we found a different proof
of Theorem~\ref{thm:flip-complex} that starts out by considering  the simplicial complex 
$\tcomplex=\tcomplex(P)$ whose faces are the sets of pairwise non-crossing edges (line segments) spanned by $P$. This complex $\tcomplex$ is shown to be a \emph{shellable simplicial ball} (by an argument based on constrained Delaunay triangulations), and $\fcomplex$ is then constructed as the \emph{dual complex} of $\tcomplex$. We hope that this alternative proof of Theorem~\ref{thm:flip-complex} is of some independent interest and present it in Sections~\ref{sec:tcomplex} and~\ref{sec:dual-complex} below.  Before that, in Section~\ref{sec:top-to-swap}, we show how to derive the Elementary Swap Theorem from Theorem~\ref{thm:flip-complex}.

\subsection{From Topology to the Elementary Swap Theorem}
\label{sec:top-to-swap}

In this section we use Theorem~\ref{thm:flip-complex} to prove the Elementary Swap Theorem.
We begin by defining elementary walks.
A \emph{walk} in the flip graph is a sequence $T_0,T_1,\dots,T_k$ of triangulations (possibly with repetitions) such that $T_{i-1}$ and $T_i$ differ by a flip. We will refer to $T_0$ and $T_k$ as the start and the end of the walk, respectively.
A walk is \emph{closed} if it starts and ends at the same triangulation.  
If $w_1$ and $w_2$ are walks such that the end of $w_1$ equals the start of $w_2$ then 
we can define their \emph{composition} $w_1w_2$ in the obvious way. 
Furthermore, if  $w=(T=T_0,T_1,\ldots,T_k)$ is a walk, we will use the notation 
$w^{-1}=(T_k,T_{k-1},\dots,T_0)$ for the \emph{inverse walk}.

Fix a triangulation $T_0$. An \emph{elementary quadrilateral walk} is a closed walk of the form 
$wzw^{-1}$, where $z$ is an elementary $4$-cycle in the flip graph, and $w$ is a walk 
from $T_0$ to some triangulation on $z$. An \emph{elementary pentagonal walk} is defined analogously, with $z$ an elementary $5$-cycle.

It is straightforward to check the effect of these elementary walks on labellings:
\begin{lemma}
\label{lem:elementary-walk-permutations} 
Let $(T_0,\ell)$ be a labelled triangulation.
An elementary quadrilateral walk does not permute the labels.
An elementary pentagonal walk swaps the labels of two edges ($e$ 
and $f$ in Figure~\ref{4-5cycle}(b))  and leaves all other labels fixed; this corresponds exactly 
to the notion of an elementary swap introduced earlier.
\end{lemma}

Another operation that does not affect the permutation of labels induced by a closed walk is the following. A \emph{spur} $ww^{-1}$ starting and ending at $T$ is an arbitrary walk $w$ starting 
at $T$, immediately followed by the \emph{inverse walk}. If $w_1$ and $w_2$ are walks in the flip graph such that $w_1$ ends at a triangulation $T$ and $w_2$ 
starts there, and if $s$ is a spur at $T$, 
then we say that the walk $w_1 s w_2$ 
differs from $w_1 w_2$ by a \emph{spur insertion}
The inverse operation is called a \emph{spur deletion}.

\begin{lemma}\label{lem-spurs}
If two closed walks $w$ and $w'$ in the flip graph differ only by a finite number of spur insertions and deletions then they yield the same permutation of edge labels.
\end{lemma}
\begin{proof}
A flip immediately followed by its inverse flip has no effect on labels. The lemma follows by induction on the length of a spur and 
the number of spur insertions and deletions.
\end{proof}

By Lemmas~\ref{lem:elementary-walk-permutations} and \ref{lem-spurs}, the Elementary Swap Theorem directly reduces to the following, which we prove using 
Theorem~\ref{thm:flip-complex}:

\begin{proposition}
\label{topological EST}
Let $w$ be a closed walk in the flip graph starting and ending at $T_0$. 
Then, 
up to a finite number of spur insertions and deletions, $w$ can be written as the composition of 
finitely many elementary walks.
\end{proposition}

\begin{proof}
We use the well-known fact that the fundamental group of a cell complex can be defined \emph{combinatorially} in terms of closed walks in the $1$-skeleton and this definition is equivalent to the usual topological definition in terms of continuous loops, see \cite[Chap.~7]{Seifert:A-Textbook-of-Topology-1980} or \cite[Chap.~4]{Stillwell:Classical-topology-and-combinatorial-group-1993}. In particular, in a cell complex with trivial fundamental group any two closed walks in the $1$-skeleton starting at the same vertex are related by a finite number of spur insertions, deletions and so-called 2-cell relations.  

We describe the combinatorial definition of the fundamental group of the flip complex $\fcomplex$ in detail. By Theorem~\ref{thm:flip-complex}, the $1$-skeleton of $\fcomplex$ is the flip graph of $P$. Fix a \emph{base triangulation} $T_0$, and, for every triangulation $T$, fix a walk $p_T$ from $T_0$ to $T$. Given two triangulations $T_1,T_2$ that differ by a flip, we form the closed walk $w_{T_1,T_2}$ in the flip graph, called a \emph{generating walk}, that goes from $T_0$ to $T_1$ along $p_{T_1}$, then flips to $T_2$, and then returns to $T_0$ along $p^{-1}_{T_2}$.  It is easy to see that, up to a finite number of spur insertions and deletions, every closed walk starting and ending at $T_0$ can be written as a composition of generating walks.

We say that walks $w$ and $w'$ are \emph{$2$-cell related} if we can express them as
$w=w_1w_2$ and $w'=w_1z w_2$, where $z$ is a closed walk traversing the boundary 
of a $2$-cell (an elementary cycle) exactly once in either orientation. 
Notice that $w_1w_2$ and $w_1zz^{-1}w_2$ differ only by the spur $zz^{-1}$, hence, up to spur insertion and deletion, being $2$-cell related is symmetric. 
 
Also, notice the \emph{precomposition property}: if $w$ and $w'$ are $2$-cell related as above and if $w$ is precomposed with the
closed walk $w_1z w_1^{-1}$ then the result $w''=(w_1z w_1^{-1})w=w_1z(w_1^{-1}w_1)w_2$ differs from $w'$ only by the spur $w_1^{-1}w_1$. By Theorem~\ref{thm:flip-complex}, a boundary of a $2$-cell is an elementary $4$- or $5$-cycle and so the walk $w_1z w_1^{-1}$ above is an elementary walk.

Two walks in the flip graph are called equivalent if they differ by a finite number of spur insertion and/or deletions and by applying a finite number of $2$-cell relations. It is not hard to check that this defines an equivalence relation, and the fundamental group $\pi_1(\fcomplex)$ is given as the set of equivalence classes of closed walks starting and ending at $T_0$. 

By Theorem~\ref{thm:flip-complex}, the fundamental group of the flip complex $\fcomplex$ is trivial. This translates into the fact that every closed walk starting and ending at $T_0$ is equivalent to the trivial walk. By 
the precomposition property, this means that, 
up to a finite number of spur insertions and deletions, every closed walk is a composition of finitely many elementary walks.
\end{proof}

\subsection{The Simplicial Complex of 
Plane Graphs} 
\label{sec:tcomplex}

In this section and the following one, we give a proof of Theorem~\ref{thm:flip-complex}.
This section is about the simplicial complex 
$\tcomplex=\tcomplex(P)$ whose faces are the sets of pairwise non-crossing edges (line segments) spanned by $P$.

Let $P$ be a set of $n$ points in general position in the plane.
Let $E$ be the set of edges (closed line segments) spanned by $P$.
Two edges $e,f\in E$ are said to be \emph{non-crossing} if they
are disjoint or if they intersect in a single point of $P$ that is an endpoint of both
edges. We say that a subset $F\subseteq E$ is \emph{non-crossing}
if every pair of distinct edges $e,f\in F$ is non-crossing. If $G$ is non-crossing and $F\subseteq G$ then $F$ is non-crossing as well. Thus, the non-crossing sets
of edges form an abstract simplicial complex

$$\tcomplex=\tcomplex(P):=\{F \colon F\subseteq E, F \textrm{ non-crossing}\},$$
which we call the \emph{complex of 
plane graphs on $P$}. 
We collect some basic properties of $\tcomplex$:

\begin{enumerate}
\item The \emph{facets} (inclusion-maximal faces) of $\tcomplex$ are exactly the triangulations of $P$
(every non-crossing set of edges $F\subseteq E$ can be extended to a 
triangulation). Thus, the simplicial complex $\tcomplex$ is of dimension $m-1$,
where $m$ is the number of edges in any triangulation of $P$, 
and it is \emph{pure}, i.e., every face of $\tcomplex$ is
contained in a face of dimension $m-1$.
\item Every face $F$ of $\tcomplex$ of dimension $m-2$ is contained in either one or two triangulations. In the latter case, $F$ corresponds to a flip between these two triangulations.
\end{enumerate}

We will show that the topology of $\tcomplex$ is particularly simple, namely that $\tcomplex$ 
is a homeomorphic to an $(m-1)$-dimensional ball. Furthermore, there is a combinatorial certificate 
(\emph{shellability}) for this homeomorphism. This implies that the homeomorphism is particularly nice 
and that $\tcomplex$ is a \emph{piecewise-linear ball}. We refer to  \cite{Hudson:Piecewise-linear-topology-1969} and \cite[Appendix~4.7]{Bjorner:Oriented-matroids-1999} for more details and further references on shellability and piecewise-linear balls, spheres, and manifolds.
In this extended abstract, we will leave the notion of piecewise-linearity undefined---the only property that we will need is that it ensures that the construction of the \emph{dual cell complex} $\tcomplex^*$ (see Proposition~\ref{prop-dual} below) is well-behaved.

We recall that a pure $d$-dimensional simplicial complex is \emph{shellable} if 
there exists a total ordering of its facets $F_{1},F_{2},\cdots,F_{N}$ (called a \emph{shelling order})
such that, for every $2\leq j\leq N$, the intersection of $F_j$ with the simplicial complex 
generated by the preceding facets\footnote{More formally, for any set $F$, let $2^F$ denote the simplicial complex of all subsets of $F$. Then the requirement for a shelling is that, for $2\leq j\leq N$, the intersection of the complexes $2^{F_j}$ and $\bigcup_{i<j}2^{F_i}$ be pure of dimension $d-1$.} is pure of dimension $d-1$.

We will need the following result (which appears implicitly in \cite{Bing:Some-aspects-of-the-topology-of-3-manifolds-1964}, and explicitly in \cite{danaraj1974shellings}; see \cite[Prop.~4.7.22]{Bjorner:Oriented-matroids-1999} for a short proof):
\begin{proposition}
\label{prop:shellable}
Suppose $\mathbbm{K}$ is a finite $d$-dimensional simplicial complex that is a pseudomanifold, i.e., $\mathbbm{K}$ is pure and every $(d-1)$-dimensional face of $\mathbbm{K}$ is contained in at most two $d$-faces. 
If $\mathbbm{K}$ is shellable then $\mathbbm{K}$ is either a piecewise-linear ball or a piecewise-linear sphere. The former case occurs iff 
there is at least one $(d-1)$-dimensional face that is contained in only one $d$-face of $\mathbbm{K}$.\footnote{We remark that the property of being a shellable pseudomanifold (which is a combinatorial and algorithmically verifiable condition) is strictly stronger than being a piecewise-linear ball or sphere, which in turn is strictly stronger than being a simplicial complex homeomorphic to a ball or sphere.}
\end{proposition}

\begin{theorem}
\label{Ball theorem} 
\label{ball thm}
$\tcomplex$ is \emph{shellable}, and hence a piecewise-linear $(m-1)$-dimensional ball.
\end{theorem}

\begin{proof}
We observed earlier that $\tcomplex$ is a pure $(m-1)$-dimensional simplicial complex,
and that every $(m-2)$-dimensional face of $\tcomplex$ is contained in at most two $(m-1)$-dimensional faces, hence $\tcomplex$ is a pseudomanifold. Moreover, if $T$ is a triangulation of $P$ and if $e\in T$ is a non-flippable edge
(e.g., if $e$ is a convex hull edge) then $F:=T\setminus\{e\}$ is an $(m-2)$-dimensional face of $\tcomplex$ that is contained in a unique $(m-1)$-face, namely $T$.

Thus, by Proposition~\ref{prop:shellable}, it suffices to show that $\tcomplex$ is shellable, i.e., to exhibit a shelling order for the facets of $\tcomplex$. 

With every triangulation $T$ of $P$, we associate the sorted vector of angles $\alpha(T)=(\alpha_{1}(T),\alpha_{2}(T),\cdots,\alpha_{3t}(T))$, where $\alpha_{1}(T) \leq \alpha_{2}(T) \leq \cdots \leq\alpha_{3t}(T)$ are the angles occurring in the triangulation $T$. We order the triangulations of $P$ 
by sorting the corresponding angle vectors $\alpha(T)$ lexicographically from largest to smallest; if the point set is in general position, this defines a total ordering
\begin{equation}
\label{eq:angle-ordering}
T_1,T_2,\dots,T_N,\qquad \alpha(T_1)>_{\textrm{LEX}}\alpha(T_2)>_{\textrm{LEX}}\dots>_{\textrm{LEX}}\alpha(T_N),
\end{equation}
where $N$ is the number of triangulations of $P$.

It is well known (see, for example, \cite[Chap.~3.4]{devadoss2011discrete})
that in this ordering, $T_{1}$ is the Delaunay triangulation of
$P$. Moreover, if we consider only triangulations containing a particular
plane subgraph corresponding to a face $F$ of $\tcomplex$ 
and the corresponding subsequence of the angle vectors,
the first of these vectors corresponds to the Delaunay triangulation
constrained to $F$. 

We claim that the triangulation ordering \eqref{eq:angle-ordering} defines a
shelling.  For this, we need to prove that the following holds for $2\leq j\leq N$:
If $F$ is a face of $\tcomplex$ that is contained in $T_j\cap T_i$ for some $i<j$,
then there exists an $(m-2)$-dimensional face $G$ of $T_j$ and some $i'<j$
such that $F\subseteq G=T_{i'}\cap T_j$.

To see this, consider the subsequence 
$T_{k_1},T_{k_2},\dots $ of the sequence \eqref{eq:angle-ordering} consisting only of those triangulations that contain the edge set $F$. Then $T_{k_1}$ is the constrained Delaunay 
triangulation with respect to the edge set $F$, and $T_i$ and $T_j$ both appear in that
subsequence; in particular, $T_j\neq T_{k_1}$ since $T_i$ precedes it. Since every triangulation
containing $F$ can be transformed to the constrained Delaunay triangulation $T_{k_1}$,
(see, e.g., the description of the Lawson flip algorithm in \cite{devadoss2011discrete})
there must exist an edge $e\in T_{j}\setminus T_{k_1}$ such that flipping $e$ (a Lawson flip)
increases the angle vector; thus, the triangulation resulting from flipping $e$ is some
$T_k$ with $k<j$ and satisfies $F\subseteq T_k\cap T_j$ as desired.
\end{proof}

Finally, we need a characterization of interior versus boundary faces of  $\tcomplex$.
Let $\mathbbm{B}$ be a piecewise-linear ball of dimension $d$. By definition, the \emph{boundary} $\partial\mathbbm{B}$ of $\mathbbm{B}$ is the subcomplex of $\mathbbm{B}$ consisting of all faces $F$ for which there exists a $(d-1)$-dimensional face $G$ of 
$\mathbbm{B}$, with $F\subseteq G$, such that $G$ is contained in a unique $d$-dimensional face of $\mathbbm{B}$. (In the case $\mathbbm{B}=\tcomplex$, the latter condition means that $G=T\setminus \{e\}$ for some triangulation $T$ and some edge $e\in T$ that is not flippable.)
A face $F$ of $\mathbbm{B}$ that does not lie in $\partial \mathbbm{B}$ is called an \emph{interior face}.

For the proof of Theorem~\ref{thm:flip-complex} we need properties of interior faces of $\tcomplex$ of dimensions $m-1$, $m-2$ and $m-3$.  The following proposition characterizes interior faces more generally.

\begin{proposition} 
\label{prop:interior-faces}
Let $\tcomplex$ be the simplicial complex of plane graphs on the
point set $P$. A non-crossing set of edges $F$ on $P$ is an interior face of
$\tcomplex$ if and only if the following conditions hold:

(i) $F$ contains all convex hull edges of $P$,

(ii) Every bounded region in the complement of the plane graph $\left(P,F\right)$ is convex.
\end{proposition}

\begin{proof}
Note that a polygon is non-convex iff it has a reflex vertex.  More generally,
a bounded region in the complement of the plane graph $\left(P,F\right)$ is non-convex iff there is an interior point $p$ of $P$ and a half-plane $H$ through $p$ with no edge of $F$ from $p$ to a point interior to $H$---in this case we say that $p$ ``has no edge in a half-plane''.  
The statement of the proposition is then equivalent to the following:
$F$ is a boundary face if and only if $F$ misses a convex hull edge or there is an interior point $p$ of $P$
with no edge in a half-plane.  We prove this statement.  

For the forward direction, 
suppose that $F$ is a boundary face.  Then there is a triangulation $T$, $F \subseteq T$, and an edge $e \in T - F$ such that $e$ is not flippable in $T$.  If $e$ is a convex hull edge, then $F$ does not contain all convex hull edges.  Otherwise $e$ is a diagonal of a non-convex quadrilateral in $T$.  Set $p$ to be the reflex vertex of the non-convex quadrilateral and $H$ to contain the other end of $e$ but not the two other vertices of the quadrilateral.  Then $p$ has no edge in half-plane $H$.

For the other direction, first note that if 
$F$ misses a convex hull edge then $F$ is a boundary face.  For the other case, suppose there is a non-convex hull point $p$ of $P$ that has no edge in half-plane $H$.  
Augment $F$ to a maximal set $F'$ of non-crossing edges without using any edge from $p$ into $H$.  This will not yet be a triangulation (because in a triangulation $p$ is surrounded by triangles and they have angles bounded by $\pi$).  Now augment further to a triangulation $T$.  Then $T - F'$ contains some edge $e$ incident to $p$, and $e$ is not flippable otherwise we could have further augmented $F'$. Thus $F$ is a boundary face.
\end{proof}

\subsection{The Dual Flip Complex $\fcomplex$}
\label{sec:dual-complex}

To define the flip complex $\fcomplex$, we need the notion of \emph{dual cells} and the dual cell decomposition of a piecewise-linear ball; for the precise definition, we refer to \cite[Sec.~I.6]{Hudson:Piecewise-linear-topology-1969} or \cite[\S64 and \S70]{Munkres:Elements-of-algebraic-topology-1984}.\footnote{In \cite{Munkres:Elements-of-algebraic-topology-1984}, the terminology \emph{dual blocks} is used instead of dual cells, since the construction is described in a more general setting (for arbitrary triangulated manifolds or homology manifolds) in which the dual 
blocks need not be cells (homeomorphic to balls). In the setting of piecewise-linear manifolds, in particular piecewise-linear balls, however, this technical issue does not arise.}
Here, we simply collect the properties that we will need:

\begin{proposition} 
\label{prop-dual}
Let $\mathbbm{B}$ be a $d$-dimensional piecewise-linear ball.
\begin{enumerate}
\item For each interior $k$-dimensional face $F$ of $\mathbbm{B}$, one can define a \emph{dual cell} $F^*$ (a certain subcomplex of the \emph{barycentric subdivision} of 
$\mathbbm{B}$ that is a piecewise-linear ball of dimension $d-k$ \cite[Lemma~I.19]{Hudson:Piecewise-linear-topology-1969}).
\item The construction reverses inclusion, i.e., 
for interior faces $F$, $G$ of $\mathbbm{B}$,
$F\subseteq G$ iff $F^* \supseteq G^*$.
\item The dual cells of the 
interior faces of $\mathbbm{B}$ form a regular cell complex, denoted $\mathbbm{B}^*$ and called the \emph{dual cell complex}.
$\mathbbm{B}^*$ need not be a manifold or pure $d$-dimensional, but it is homotopy equivalent to $\mathbbm{B}$ \cite[Lem.~70.1]{Munkres:Elements-of-algebraic-topology-1984}.\footnote{More specifically, the dual complex of a piecewise-linear manifold with boundary is a deformation retraction of the manifold. For manifolds without boundary, the dual complex is piecewise-linearly homeomorphic to the original manifold.}
\end{enumerate}
\end{proposition}

We define the \emph{flip complex} $\fcomplex:=\tcomplex^*$ 
as the dual complex of the simplicial complex  $\tcomplex$.

\begin{proof}[Proof of Theorem~\ref{thm:flip-complex}] 
By Proposition~\ref{prop-dual}, $\fcomplex=\tcomplex^*$ is a regular cell complex that is homotopy equivalent to the ball 
$\tcomplex$; consequently, the fundamental group $\pi_1(\fcomplex)$ vanishes. 

It remains to show the characterization of the vertices, edges, and $2$-cells of $\fcomplex$.

The vertices of $\fcomplex$ correspond (are dual) 
to the faces of $\tcomplex$ of the highest dimension $(m-1)=\dim \tcomplex$, i.e., to the triangulations of $P$ (these are automatically interior faces of $\tcomplex$). 

The edges of $\fcomplex$ correspond to interior $(m-2)$-dimensional faces $F$ of $\tcomplex$,
i.e., faces $F$ that are contained in two triangulations $T$ and $T'$ that differ by a flip.
Thus, the $1$-skeleton of $\fcomplex$ is exactly the flip graph of $P$.

Every $2$-cell of $\fcomplex$ is the dual cell $F^*$ of an interior face $F$ of 
$\tcomplex$ of dimension $m-3 = \dim F$.
Consider an arbitrary triangulation $T$ containing $F$, i.e., $F$ is obtained
from $T$ by deleting two edges $e,f$. 
By Proposition~\ref{prop:interior-faces}, $e$ and $f$ are both flippable in $T$ since they lie in a convex polygon in $T$.

If $e$ and $f$ are not incident to a common triangle in $T$, (or, equivalently, removing both $e$ and $f$ from $T$ creates two internally disjoint convex quadrilaterals) then there exist four triangulations containing $F$
and these form an elementary $4$-cycle in the flip graph. The $4$-cycle is by definition the boundary of the 
dual cell $F^*$. 

Otherwise, $e$ and $f$ are incident to a common triangle in $T$.  By Proposition~\ref{prop:interior-faces} the union of the three triangles of $T$ containing either $e$ or $f$ forms a convex polygon, necessarily a pentagon.
There are five 
triangulations containing $F$ and these form an elementary $5$-cycle in the flip graph. 
The 5-cycle is by definition the boundary of the dual cell $F^*$.

%
%

Hence, every $2$-cell of $\fcomplex$ corresponds to an elementary $4$- or $5$-cycle of the flip graph.

Conversely, every elementary $4$- or $5$-cycle of the flip graph gives rise to a 2-cell $F^*$ of $\fcomplex$: more precisely, $F^*$ corresponds to the intersection of the triangulations in the elementary cycle.  
\end{proof}



\section{Proofs of Properties of Elementary Swaps}
\label{sec:bounds}

In this section we prove Lemmas~\ref{lemma:elem-swap} and \ref{lemma:elem-swap-seq}.

To prove 
Lemma~\ref{lemma:elem-swap}, the idea is to look at paths in the \textit{double quadrilateral graph} $G_D$ that we will define below. Informally speaking, 
$G_D$ captures where pairs of non-crossing edges can go via flips, similar to the way the quadrilateral graph captures where a single edge can go via flips.
We will show that there is an elementary swap between two labels in a triangulation if and only if there exists a path of certain type in the double quadrilateral graph. 

\begin{proof}[Proof of Lemma~\ref{lemma:elem-swap}]
Construct a graph $G_D$ called the \emph{double quadrilateral graph}. 
Vertices of the graph $G_D$ are
pairs of non-crossing edges on the point set $P$, and we define two vertices
$(e_{1},f_{1})$ and $(e_{2},f_{2})$ of $G_D$ to be adjacent
if either $e_{1}=e_{2}$ and $f_{1}$ and $f_{2}$
are adjacent in the quadrilateral graph, 
or if $f_{1}=f_{2}$ and $e_{1}$ and $e_{2}$
are adjacent in the quadrilateral graph.
(Recall that two edges $a$ and $b$ are adjacent in the quadrilateral graph if $a$ and $b$ cross and their four endpoints form an empty quadrilateral.)

In the graph $G_D$ we identify some vertices as ``swap vertices''.  These are the vertices $(g,h)$ such that $g$ and $h$ are diagonals of some empty convex pentagon in the point set. 
Note that the swap vertices can be identified in polynomial time.

We claim that  there is an elementary swap of $e$ and $f$ in labelled triangulation $\cal T = (T,\ell)$  if and only if there is a path in $G_D$ from vertex $(e,f)$ to a swap vertex.  
For the forward direction, suppose there is such an elementary swap.
It begins with a sequence $\sigma$ of flips from $\cal T$ to a labelled triangulation ${\cal T}'$ in which
labels $\ell(e)$ and $\ell(f)$ are attached to two diagonals $g$ and $h$ of some empty convex pentagon.
The subsequence of $\sigma$ consisting of those flips that apply to an edge whose current label is $\ell(e)$ or $\ell(f)$ corresponds to a path in $G_D$ from $(e,f)$ to the swap vertex $(g,h)$.

For the other direction, let $\pi$ be a path in $G_D$ from $(e,f)$ to a swap vertex.  
It suffices to show that the path $\pi$ provides a sequence of flips, $\sigma$, that takes $\cal T$ to some labelled triangulation ${\cal T}'$ in which
labels $\ell(e)$ and $\ell(f)$ are attached to two diagonals of an empty convex pentagon,
because the rest of the elementary swap is then determined.  
Consider the first edge of $\pi$ and suppose without loss of generality that it goes from $(e,f)$ to $(e,f')$ (the case when $e$ changes is similar).   Then $e$ and $f'$ are non-crossing.  Because $f$ and $f'$ are adjacent in the quadrilateral graph, they cross and form an empty convex quadrilateral $Q$.
Note that $e$ does not intersect the interior of $Q$, since $Q$ is empty and $e$ does not cross $f$ or $f'$.
We apply the result that any constrained triangulation can be  flipped to any other with $O(n^2)$ flips.  Fix edges $e$ and $f$ in $T$ and flip $\cal T$ to a labelled triangulation that contains the edges of $Q$.  In this triangulation, we can flip $f$ to $f'$, transferring $\ell(f)$ to $f'$.  We continue in this way to realize each edge of $\pi$ via $O(n^2)$ flips, arriving finally at a labelled triangulation in which labels $\ell(e)$ and $\ell(f)$ are attached to edges that are the diagonals of some empty convex pentagon in the point set.  Fixing the two diagonals, we can flip to a triangulation that contains the edges of the convex pentagon, and at this point we are done.      

Because the graph $G_D$ has $O(n^4)$ vertices, the diameter of any of its connected components is $O(n^4)$.
Thus, if there is an elementary swap that exchanges the labels of edges $e$ and $f$, then there is one corresponding to a path in $G_D$ of length $O(n^4)$.
We can explicitly construct $G_D$ and find such a path in polynomial time.
As argued above, every edge of $G_D$ can be realized by $O(n^2)$ flips.
This proves that, for any elementary swap, we can construct a sequence of $O(n^6)$ flips to realize it, and the construction takes polynomial time.  
\end{proof}


As  
mentioned in Section~\ref{sec:reductions}, there is a group-theoretic argument proving a weaker version of Lemma~\ref{lemma:elem-swap-seq}.  The argument depends on the following claim: If a permutation group is generated by transpositions and contains a permutation that maps element $e$ to $f$ then the group contains the transposition of $e$ and $f$. To prove this claim, notice that if the group contains transpositions $(ab)$ and $(bc)$, then it also contains transposition $(ac)=(ab)(bc)(ab)$; and apply induction.

To apply this claim in our situation, observe that
by the Elementary Swap Theorem, all label permutations achievable by flips in a triangulation $\cal T$ are compositions of elementary swaps, hence, these label permutations indeed form a group $G$ generated by transpositions. Moreover, by the assumption of Lemma~\ref{lemma:elem-swap-seq}, $G$ contains a permutation taking the label of edge $e$ to edge $f$. Hence, by the above claim, the group $G$ also contains a 
permutation, which is a composition of elementary swaps, whose effect is to transpose labels of edges $e$ and $f$.

In order to prove the full result of Lemma~\ref{lemma:elem-swap-seq}, i.e., that the label transposition of $e$ and $f$ can be done with a single elementary swap, we combine the techniques used in the proof of the group theory claim above with the structure of elementary swaps. 

\begin{proof}[Proof of Lemma~\ref{lemma:elem-swap-seq}]
An elementary swap in triangulation $\cal T$ acts on two edges of $\cal T$.  
We define a graph $G_S$ called the \emph{elementary swap graph} of $\cal T$. 
$G_S$ has a vertex for every edge of $\cal T$, and we define vertices $e$ and $f$ to be adjacent in $G_S$ if there is an elementary swap  of $e$ and $f$ in $\cal T$. 

By hypothesis, there is a sequence of elementary swaps that takes the label of edge $e$ to edge $f$.
Observe that no sequence of elementary swaps will take the label of edge $e$ outside the connected component of $G_S$ that contains $e$.
Therefore $e$ and $f$ must lie in the same connected component of $G_S$.  
We will now show that 
each connected component of $G_S$ is a clique.  This implies that there is an 
elementary swap of $e$ and $f$, and completes our proof.

Consider a simple path $(e_0,e_1), (e_1, e_2), \ldots, (e_{k-1}, e_k)$ in $G_S$.  
Let $\sigma_i$, $i=1, \ldots, k$ be a flip sequence that realizes the elementary swap $(e_{i-1}, e_i)$, and let $\sigma = \sigma_1 \sigma_2 \ldots \sigma_{k-1}$.  Observe that $\sigma$ takes the label of $e_0$ to $e_{k-1}$, and does not change the label of $e_k$ (by the assumption that the path is simple).
By definition of an elementary swap, the flip sequence $\sigma_k$ has the form $\rho \pi \rho^{-1}$ where $\rho$ is a sequence of flips that moves the labels of $e_{k-1}$ and $e_k$ into an empty convex pentagon, and $\pi$ is the sequence of five flips that exchanges the labels of $e_{k-1}$ and $e_k$. 

Consider the flip sequence  $\sigma \sigma_k \sigma^{-1} = \sigma \rho \pi \rho^{-1} \sigma^{-1} = \sigma \rho \pi (\sigma \rho)^{-1}$. 
The first part of this flip sequence, $\sigma \rho$, moves the labels of $e_0$ and $e_k$ into an empty convex pentagon; the middle part, $\pi$, exchanges them; and the final part, $(\sigma \rho)^{-1}$ reverses the first part.  Therefore this flip sequence realizes an elementary swap of $e_0$   
and $e_k$.  
\end{proof}


\section{Conclusions}
\label{sec:conclusions}

We have characterized when 
two labelled triangulations of a set of $n$ points belong to the same connected component of the labelled flip graph, and proved that the diameter of each connected component is bounded by $O(n^7)$.
We conclude with some open problems:

\begin{enumerate}

\item 
Reduce the gap between the upper bound, $O(n^7)$, and the best known lower bound of $O(n^3)$~\cite{bose2013flipping} on the diameter of a component of the labelled flip graph.

\item 
We have studied the case where each edge in a triangulation has a unique label, and given a bound of $O(n^7)$ on the diameter of a component of the labelled flip graph.  The case where edges are unlabelled can be viewed as the case where every edge has the same label---in this case the bound becomes $O(n^2)$.  
A unifying scenario is when the edges have labels and labels may appear on more than one edge.
Is there a bound on the diameter of connected components of the flip graph that depends on the number of labels, or on the maximum number of edges with the same label?
%

\item
We did not analyze the run-time of our algorithms in the main text.  A crude bound is $O(n^8)$, with the bottleneck being the explicit construction in the proof of Lemma~\ref{lemma:elem-swap} of the double quadrilateral graph which has $O(n^4)$ vertices and thus $O(n^8)$ edges.  This bound can surely be improved. 

 \item 
 What is the complexity of the following flip distance problem for labelled triangulations: Given two labelled triangulations and a number $k$, is there a flip sequence of length at most $k$ to transform the first triangulation to the second one?
This problem is NP-complete in the unlabelled setting, but 
knowing the mapping of edges might make the problem easier.

\end{enumerate}

\subparagraph*{Acknowledgements}
This research was initiated at the 2016 Bellairs Workshop on Geometry and Graphs. We thank anonymous reviewers and participants of the 2017 Symposium on Computational Geometry for helpful suggestions.

\bibliography{references}

\begin{thebibliography}{10}

\bibitem{aichholzer2015flip}
Oswin Aichholzer, Wolfgang Mulzer, and Alexander Pilz.
\newblock Flip distance between triangulations of a simple polygon is
  {NP}-complete.
\newblock {\em Discrete \& Computational Geometry}, 54(2):368--389, 2015.
\newblock \href {http://dx.doi.org/10.1007/s00454-015-9709-7}
  {\path{doi:10.1007/s00454-015-9709-7}}.

\bibitem{AHOS14}
Gabriela Araujo-Pardo, Isabel Hubard, Deborah Oliveros, and Egon Schulte.
\newblock Colorful associahedra and cyclohedra.
\newblock {\em Journal of Combinatorial Theory, Series A}, 129:122--141, 2015.
\newblock \href {http://dx.doi.org/10.1016/j.jcta.2014.09.001}
  {\path{doi:10.1016/j.jcta.2014.09.001}}.

\bibitem{Bern-Eppstein}
Marshall Bern and David Eppstein.
\newblock Mesh generation and optimal triangulation.
\newblock In Ding-Zhu Du and Frank Hwang, editors, {\em Computing in
  {E}uclidean geometry}, volume~1 of {\em Lecture Notes Series on Computing},
  pages 23--90. World Scientific, 1992.
\newblock \href {http://dx.doi.org/10.1142/9789814355858_0002}
  {\path{doi:10.1142/9789814355858_0002}}.

\bibitem{Bing:Some-aspects-of-the-topology-of-3-manifolds-1964}
R.~H. Bing.
\newblock Some aspects of the topology of {$3$}-manifolds related to the
  {P}oincar{\'e} conjecture.
\newblock In {\em Lectures on modern mathematics, {V}ol. {II}}, pages 93--128.
  Wiley, New York, 1964.

\bibitem{Bjorner:Oriented-matroids-1999}
Anders Bj{{\"o}}rner, Michel Las~Vergnas, Bernd Sturmfels, Neil White, and
  G{{\"u}}nter~M. Ziegler.
\newblock {\em Oriented Matroids}, volume~46 of {\em Encyclopedia of
  Mathematics and its Applications}.
\newblock Cambridge University Press, Cambridge, 2nd edition, 1999.
\newblock \href {http://dx.doi.org/10.1017/CBO9780511586507}
  {\path{doi:10.1017/CBO9780511586507}}.

\bibitem{BH09}
Prosenjit Bose and Ferran Hurtado.
\newblock Flips in planar graphs.
\newblock {\em Computational Geometry Theory and Applications}, 42(1):60--80,
  2009.
\newblock \href {http://dx.doi.org/10.1016/j.comgeo.2008.04.001}
  {\path{doi:10.1016/j.comgeo.2008.04.001}}.

\bibitem{bose2013flipping}
Prosenjit Bose, Anna Lubiw, Vinayak Pathak, and Sander Verdonschot.
\newblock Flipping edge-labelled triangulations.
\newblock {\em arXiv:1310.1166}, 2013.
\newblock To appear in {\it Computational Geometry}.
\newblock URL: \url{http://arxiv.org/abs/1310.1166}.

\bibitem{bose2015flips}
Prosenjit Bose and Sander Verdonschot.
\newblock Flips in edge-labelled pseudo-triangulations.
\newblock {\em Computational Geometry}, 60:45--54, 2017.

\bibitem{JDHU13}
Javier Cano, Jos{\'e}-Miguel D{\'{\i}}az-B{\'a}{\~n}ez, Clemens Huemer, and
  Jorge Urrutia.
\newblock The edge rotation graph.
\newblock {\em Graphs and Combinatorics}, 29(5):1207--1219, 2013.
\newblock \href {http://dx.doi.org/10.1007/s00373-012-1201-z}
  {\path{doi:10.1007/s00373-012-1201-z}}.

\bibitem{danaraj1974shellings}
Gopal Danaraj and Victor Klee.
\newblock Shellings of spheres and polytopes.
\newblock {\em Duke Mathematical Journal}, 41(2):443--451, 1974.

\bibitem{devadoss2011discrete}
Satyan~L. Devadoss and Joseph O'Rourke.
\newblock {\em Discrete and Computational Geometry}.
\newblock Princeton University Press, 2011.

\bibitem{DGR93}
N.~Dyn, I.~Goren, and S.~Rippa.
\newblock Transforming triangulations in polygonal domains.
\newblock {\em Computer Aided Geometric Design}, 10:531--536, 1993.

\bibitem{Edelsbrunner}
Herbert Edelsbrunner.
\newblock {\em Geometry and Topology for Mesh Generation}.
\newblock Cambridge University Press, Cambridge, 2001.
\newblock \href {http://dx.doi.org/10.1017/CBO9780511530067}
  {\path{doi:10.1017/CBO9780511530067}}.

\bibitem{Eppstein}
David Eppstein.
\newblock Happy endings for flip graphs.
\newblock {\em Journal of Computational Geometry}, 1(1):3--28, 2010.
\newblock \href {http://dx.doi.org/10.20382/jocg.v1i1a2}
  {\path{doi:10.20382/jocg.v1i1a2}}.

\bibitem{Hudson:Piecewise-linear-topology-1969}
J.~F.~P. Hudson.
\newblock {\em Piecewise Linear Topology}.
\newblock W. A. Benjamin, Inc., New York-Amsterdam, 1969.

\bibitem{HNU99}
Ferran Hurtado, Marc Noy, and Jorge Urrutia.
\newblock Flipping edges in triangulations.
\newblock {\em Discrete \& Computational Geometry}, 22(3):333--346, 1999.
\newblock \href {http://dx.doi.org/10.1007/PL00009464}
  {\path{doi:10.1007/PL00009464}}.

\bibitem{IDHPSUU11}
Takehiro Ito, Erik~D. Demaine, Nicholas J.~A. Harvey, Christos~H.
  Papadimitriou, Martha Sideri, Ryuhei Uehara, and Yushi Uno.
\newblock On the complexity of reconfiguration problems.
\newblock {\em Theoretical Computer Science}, 412(12--14):1054--1065, 2011.
\newblock \href {http://dx.doi.org/10.1016/j.tcs.2010.12.005}
  {\path{doi:10.1016/j.tcs.2010.12.005}}.

\bibitem{kanj-2017}
Iyad Kanj, Eric Sedgwick, and Ge~Xia.
\newblock Computing the flip distance between triangulations.
\newblock {\em Discrete \& Computational Geometry}, 58(2):313--344, 2017.
\newblock \href {http://dx.doi.org/10.1007/s00454-017-9867-x}
  {\path{doi:10.1007/s00454-017-9867-x}}.

\bibitem{Lawson-72}
Charles~L. Lawson.
\newblock Transforming triangulations.
\newblock {\em Discrete Mathematics}, 3(4):365--372, 1972.

\bibitem{Lawson-77}
Charles~L. Lawson.
\newblock Software for {$C^1$} surface interpolation.
\newblock In {\em Mathematical Software III}, pages 161--194. Academic Press,
  New York, 1977.

\bibitem{lubiw2015flip}
Anna Lubiw and Vinayak Pathak.
\newblock Flip distance between two triangulations of a point set is
  {NP}-complete.
\newblock {\em Computational Geometry}, 49:17--23, 2015.
\newblock \href {http://dx.doi.org/10.1016/j.comgeo.2014.11.001}
  {\path{doi:10.1016/j.comgeo.2014.11.001}}.

\bibitem{lubiw-pathak-matroids}
Anna Lubiw and Vinayak Pathak.
\newblock Reconfiguring ordered bases of a matroid.
\newblock {\em arXiv:1612.00958}, 2016.

\bibitem{molloy1999mixing}
Michael Molloy, Bruce Reed, and William Steiger.
\newblock On the mixing rate of the triangulation walk.
\newblock In {\em DIMACS-AMS volume on Randomization Methods in Algorithm
  Design}, volume~43 of {\em DIMACS Series in Discrete Mathematics and
  Theoretical Computer Science}, pages 179--190. AMS, 1999.

\bibitem{Munkres:Elements-of-algebraic-topology-1984}
James~R. Munkres.
\newblock {\em Elements of Algebraic Topology}.
\newblock Addison-Wesley Publishing Company, Menlo Park, CA, 1984.

\bibitem{Orden:The-polytope-of-non-crossing-graphs-on-a-planar-2005}
David Orden and Francisco Santos.
\newblock The polytope of non-crossing graphs on a planar point set.
\newblock {\em Discrete \& Computational Geometry}, 33(2):275--305, 2005.
\newblock \href {http://dx.doi.org/10.1007/s00454-004-1143-1}
  {\path{doi:10.1007/s00454-004-1143-1}}.

\bibitem{pilz2014flip}
Alexander Pilz.
\newblock Flip distance between triangulations of a planar point set is
  {APX}-hard.
\newblock {\em Computational Geometry}, 47(5):589--604, 2014.
\newblock \href {http://dx.doi.org/10.1016/j.comgeo.2014.01.001}
  {\path{doi:10.1016/j.comgeo.2014.01.001}}.

\bibitem{Pournin13}
Lionel Pournin.
\newblock The diameter of associahedra.
\newblock {\em Advances in Mathematics}, 259:13--42, 2014.
\newblock \href {http://dx.doi.org/10.1016/j.aim.2014.02.035}
  {\path{doi:10.1016/j.aim.2014.02.035}}.

\bibitem{Seifert:A-Textbook-of-Topology-1980}
Herbert Seifert and William Threlfall.
\newblock {\em A Textbook of Topology}, volume~89 of {\em Pure and Applied
  Mathematics}.
\newblock Academic Press, 1980.

\bibitem{STT88}
Daniel~D. Sleator, Robert~E. Tarjan, and William~P. Thurston.
\newblock Rotation distance, triangulations, and hyperbolic geometry.
\newblock {\em Journal of the American Mathematical Society}, 1(3):647--681,
  1988.
\newblock \href {http://dx.doi.org/10.2307/1990951}
  {\path{doi:10.2307/1990951}}.

\bibitem{Stillwell:Classical-topology-and-combinatorial-group-1993}
John Stillwell.
\newblock {\em Classical Topology and Combinatorial Group Theory}, volume~72 of
  {\em Graduate Texts in Mathematics}.
\newblock Springer-Verlag, 2nd edition, 1993.
\newblock \href {http://dx.doi.org/10.1007/978-1-4612-4372-4}
  {\path{doi:10.1007/978-1-4612-4372-4}}.

\bibitem{Heu13}
Jan van~den Heuvel.
\newblock The complexity of change.
\newblock {\em Surveys in Combinatorics}, 409:127--160, 2013.

\end{thebibliography}
\newpage

\end{document}